\documentclass[aps, pra, reprint, twocolumn, 10pt, superscriptaddress]{revtex4-1}

\usepackage{amsmath, amssymb}
\usepackage{mathtools}
\usepackage{braket}
\usepackage{siunitx}
\usepackage{textcomp}
\usepackage{xcolor} 

\usepackage{math}
\usepackage{color}
\usepackage{hyperref}
\usepackage{braket}
\usepackage{siunitx}

\begin{document}

\title{Hyperfine interaction with the $^{229}$Th nucleus and its low lying isomeric state}

\author{Robert A. M\"uller}
\email{robert.mueller@ptb.de}
\affiliation{Physikalisch-Technische Bundesanstalt, D-38116 Braunschweig, Germany}
\affiliation{Technische Universit\"at Braunschweig, D-38106 Braunschweig, Germany}

\author{Anna V. Maiorova}
\affiliation{Center for Advanced Studies, Peter the Great St. Petersburg State Polytechnical University, Polytekhnicheskaya 29, 195251 St. Petersburg, Russia}

\author{Stephan Fritzsche}
\affiliation{Helmholtz Insitute Jena, D-07743 Jena, Germany}
\affiliation{Friedrich-Schiller-University Jena, D-07743 Jena, Germany}

\author{Andrey V. Volotka}
\affiliation{Helmholtz Insitute Jena, D-07743 Jena, Germany}

\author{Randolf Beerwerth}
\affiliation{Helmholtz Insitute Jena, D-07743 Jena, Germany}
\affiliation{Friedrich-Schiller-University Jena, D-07743 Jena, Germany}

\author{Przemyslaw Glowacki}
\altaffiliation{present address: Pozna\'n University of Technology, 60-965 Poznań, Poland}
\affiliation{Physikalisch-Technische Bundesanstalt, D-38116 Braunschweig, Germany}

\author{Johannes Thielking}
\affiliation{Physikalisch-Technische Bundesanstalt, D-38116 Braunschweig, Germany}

\author{David-Marcel Meier}
\affiliation{Physikalisch-Technische Bundesanstalt, D-38116 Braunschweig, Germany}

\author{Maksim Okhapkin}
\affiliation{Physikalisch-Technische Bundesanstalt, D-38116 Braunschweig, Germany}

\author{Ekkehard Peik}
\affiliation{Physikalisch-Technische Bundesanstalt, D-38116 Braunschweig, Germany}

\author{Andrey Surzhykov}
\affiliation{Physikalisch-Technische Bundesanstalt, D-38116 Braunschweig, Germany}
\affiliation{Technische Universit\"at Braunschweig, D-38106 Braunschweig, Germany}

\begin{abstract}
The thorium nucleus with mass number $A=229$ has attracted much interest because its extremely low lying first excited isomeric state at about $\SI{8}{eV}$ opens the possibility for the development of a nuclear clock. However, neither the exact energy of this nuclear isomer nor properties, such as nuclear magnetic dipole and electric quadrupole moment are known to a high precision so far. The latter can be determined by investigating the hyperfine structure of thorium atoms or ions. Due to its electronic structure and the long lifetime of the nuclear isomeric state, Th$^{2+}$ is especially suitable for such kind of studies. In this letter we present a combined experimental and theoretical investigation of the hyperfine structure of the $^{229}$Th$^{2+}$ ion in the nuclear ground and isomeric state. A very good agreement between theory and experiment is found for the nuclear ground state. Moreover, we use our calculations to confirm the recently presented experimental value for the nuclear magnetic dipole moment of the thorium nuclear isomer, which was in contradiction to previous theoretical studies.
\end{abstract}

\maketitle

\textit{Introduction}

While the energy levels of atomic nuclei are usually several keV, if not MeV, apart, $^{229}$Th exhibits an extremely low lying isomeric state $^{229m}$Th with an excitation energy of only about $\SI{8}{eV}$ \cite{guimaraes-filho_energy_2005, reich_energy_1990, kroger_features_1976}. Since this energy can be reached by current laser systems and the nuclear isomer is very long-lived, it has been proposed to build a nuclear clock based on the transition from the nuclear ground to the isomeric state \cite{peik_nuclear_2015}. The precision of this clock has been estimated to $10^{-19}$s \cite{campbell_single-ion_2012, kazakov_performance_2012}. Therefore such a clock may be sensitive to temporal drifts of the QCD coupling constant and the fine structure constant. In fact it was shown that the transition from the nuclear ground to the isomeric state may be orders of magnitude more sensitive to temporal variations of the fine structure constant $\alpha$ than electronic transitions \cite{flambaum_enhanced_2006, berengut_proposed_2009, thielking_laser_2017}.

The experimental realization of a nuclear clock requires a precise knowledge of the nuclear properties of both, the nuclear ground and the first excited isomeric state. Most of the important quantities are not known to a high precision so far. The exact energy of the isomer for example remains to be determined. The currently accepted value of $\SI{7.8 \pm 0.5}{eV}$ has been obtained by the comparison of fluorescence lines in the keV regime from higher excited states of the $^{229}$Th nucleus \cite{tkalya_radiative_2015, beck_energy_2007}. For the search for possible variations of fundamental constants the moments of the nuclear ground and isomeric state \cite{flambaum_enhanced_2006, berengut_proposed_2009} are of major importance. The moments of the ground-state nucleus have been extracted to a good precision by a combination of theory and experiment in Th$^{3+}$ \cite{safronova_magnetic_2013}. Moreover, recently first measurements of the moments of the nuclear isomer have been presented \cite{thielking_laser_2017}. Since these were partially contradicting previous calculations \cite{dykhne_matrix_1998, litvinova_nuclear_2009}, a theoretical explanation is still pending.

A well established and precise method for determining nuclear moments is hyperfine spectroscopy; the measurement of the hyperfine splitting of electronic levels. The precise analysis of such experimental data requires both, theoretical and experimental effort. A theoretically challenging, but experimentally convenient system for hyperfine spectroscopy of thorium is the charge state Th$^{2+}$, which resembles a two-valence electron system. The ionization threshold of this ion is well above the low lying nuclear resonance \cite{von_der_wense_direct_2016} and the density of low energetic electronic levels is much smaller than in the case of Th$^+$. Therefore the longevity of the nuclear isomer is increased, allowing for a better measurement statistic. However except for the recent experiment \cite{thielking_laser_2017}, neither experimental nor theoretical values for the hyperfine structure in $^{229}$Th$^{2+}$ have been presented so far.

In this letter we present a comparative study of experimental values and theoretical calculations for the hyperfine splitting in both $^{229}$Th$^{2+}$ and $^{229m}$Th$^{2+}$. Our theoretical results are obtained using two different methods, a combination of configuration interaction and second order many-body perturbation theory (CI+MBPT) and the multi-configurational Dirac-Fock approach (MCDF). The combination of theory and experiment allows us to refine the results for the moments of the nuclear isomer. This might support future investigation of possible variations of fundamental constants.

\textit{Hyperfine structure of thorium ions}

Hyperfine spectroscopy is the measurement of the splitting of atomic levels due to the coupling of the nuclear spin $\vec{I}$ and the total angular momentum $\vec{J}$ of the electronic state to the combined total angular momentum $\vec{F}=\vec{I}+\vec{J}$. The energy shift induced by that coupling can be calculated using first order perturbation theory
\begin{subequations}
\begin{align}
E_{M1} &= \frac{1}{2} A C,\\
E_{E2} &= B \frac{\frac{3}{4}C(C+1)-I(I+1)J(J+1)}{2I(2I-1)J(2J-1)},
\end{align}
\label{eq:hfs splitting}
\end{subequations}
where $C=F(F+1)-J(J+1)-I(I+1)$. The subscript $M1$ refers to the energy shift due to the interaction of the electron shell with the magnetic dipole moment of the nucleus, while $E2$ is the corresponding shift due to the nuclear electric quadrupole moment. Usually, higher multipoles than $E2$ interaction are negligible.

The hyperfine energies \eqref{eq:hfs splitting} scale with the so called \emph{hyperfine constants} $A$ and $B$. For an electronic state characterized by its total angular momentum $J$ and further quantum numbers $\gamma$ these constants are obtained as \cite{grant_relativistic_2007}:
\begin{subequations}
\begin{align}
A &= \frac{\mu_I}{I}\frac{1}{\sqrt{J(J+1)(2J+1)}}\braket{\gamma J||\vec{T}^1||\gamma J},\\
B &= 2Q\left[\frac{J(2J-1)}{(J+1)(2J+1)(2J+3)}\right]^{\frac{1}{2}}\braket{\gamma J||\vec{T}^2||\gamma J},
\end{align}
\label{eq:hfs constants}
\end{subequations}
It can be seen from the equation that $A$ and $B$ are proportional to the nuclear magnetic dipole and electric quadrupole moment $\mu_I$ and $Q$, respectively. The operators $\vec{T}^k$ are the electronic parts of the hyperfine interaction. Relation \eqref{eq:hfs constants} can be used to extract the nuclear moments from the hyperfine splitting of atomic lines. 

\textit{Numerical calculations}

The theoretical determination of the hyperfine constants $A$ and $B$ and, hence, the hyperfine structure of Th$^{2+}$ requires the evaluation of the many electron matrix elements of the hyperfine operator $\vec{T}^k$ (cf. Eq. \eqref{eq:hfs constants}). We therefore need to obtain a precise representation of the many-electron wave functions of the corresponding states. In this paper we apply two different methods to approximate these wave functions.

The first technique we use to calculate hyperfine constants in $^{229}$Th$^{2+}$ is relativistic configuration interaction (CI). In CI the ansatz for the many-electron wave function with well defined parity $\Pi$ and total angular momentum $J$ is a superposition of $N$ so called configuration state functions (CSFs) \cite{grant_relativistic_2007}:
\begin{equation}
\Psi(\Pi, J) = \sum_{i=1}^{N} c_i \Phi(\Pi, J).
\label{eq:wf ci}
\end{equation}
With these basis functions the hamiltonian matrix and its eigenvalues are calculated. The entries of the corresponding eigenvectors are the expansion coefficients $c_i$. This reduces the problem of constructing a complicated many-electron wave function to the diagonalization of a matrix.

The relativistic CI approach is well known for its performance to calculate QED-corrections to atomic processes up to a very high precision \cite{artemyev_qed_2007, yerokhin_two-photon_2000}. For the determination of hyperfine constants, however, the correlation between core electrons and core-valence corellations are of major importance, which are typically neglected in CI for practical reasons. To overcome these problems many-body perturbation theory (MBPT) can be used to account for core-core and core-valence correlations, while the dynamics of the valence electrons are still treated using CI. A detailed description of this method can be found e.g. in Refs. \cite{kozlov_ci-mbpt:_2015, dzuba_combination_1996, dzuba_core-valence_2007, dzuba_v_2005}.

Accurate wave functions can be constructed alternatively by applying the multi-configurational Dirac-Fock (MCDF) method \cite{grant_relativistic_2007}. Here the wave functions are also constructed as a superposition of CSFs. However, in contrast to CI the CSFs are not fixed, but iteratively optimized to achieve self-consistency of the result. For our calculations we utilized the newest version of the \textsc{grasp2k} package \cite{jonsson_new_2013}.

%

\textit{Experimental method}
\begin{figure}[tb]
\includegraphics[width=\linewidth]{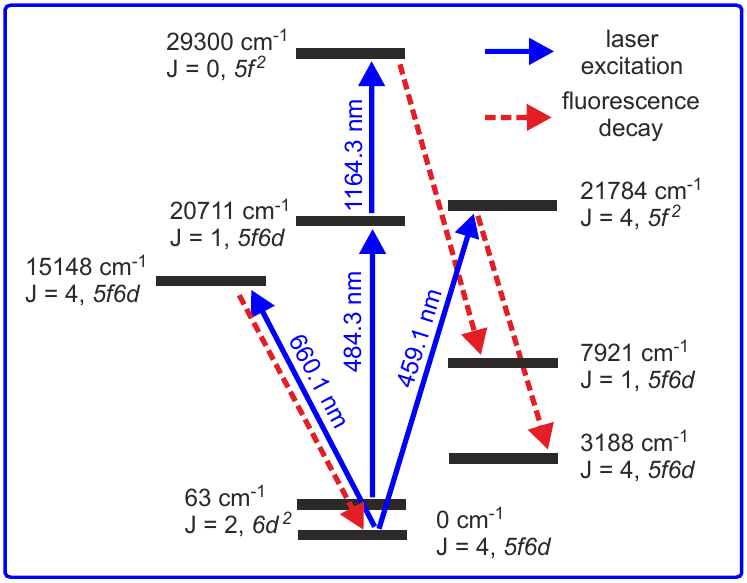}
\caption{\label{fig:levels} Scheme of the investigated $^{229}$Th$^{2+}$ levels. The transitions and electronic configurations of Th$^{2+}$ levels relevant to the experiment are depicted, labeled by their energy in cm$^{-1}$ and the electronic angular momentum $J$. Laser excitation is shown with solid arrows and fluorescence detection with dashed arrows.}
\end{figure}

For the experimental investigation of the $^{229}$Th$^{2+}$ hyperfine structure we use ions stored in a radio-frequency linear Paul trap \cite{herrera_two-photon_2012}. The ions are cooled to room temperature by collisions with a buffer gas (He) at 0.1 Pa pressure, which also depopulates metastable states via collisional quenching. To prepare a sample of Th$^{2+}$ ions we first load $10^{5}$ Th$^{+}$ ions into the trap via laser ablation and then generate doubly charged ions via photo-ionization. We reach a stable amount of $10^{3}$ Th$^{2+}$ ions, defined by an equilibrium of the ionization rate and losses due to chemical reactions with impurities in the buffer gas.

We study single-photon excitations at two different frequencies which can be addressed by external-cavity diode lasers (see Fig. \ref{fig:levels}). The levels $5f^2(J=4):\SI{15148}{cm^{-1}}$ and $5f^2(J=4):\SI{21784}{cm^{-1}}$ are excited from the electronic ground state $5f6d(J=4):\SI{0}{cm^{-1}}$ via laser radiation at 660.1 nm and 459.1 nm, respectively. Both upper levels possess a fluorescence decay channel in the visible spectrum detectable by photomultiplier tubes. In addition, the decay from $5f^2(J=4):\SI{21784}{cm^{-1}}$ is spectrally separated from the excitation, allowing for a detection free from background of laser stray light.

To achieve a higher resolution we use a two-step laser excitation, free from Doppler-broadening ~\cite{bjorkholm_line_1976, kalber_nuclear_1989}. In this case one laser excites a narrow velocity class of ions out of the thermal distribution to an intermediate state, where they are probed by resonant excitation to a higher-lying level using a second tunable laser. We have selected the transition from the $6d^2(J=2):\SI{63}{cm^{-1}}$ electronic state to $5f^2(J=0):\SI{29300}{cm^{-1}}$ via the $5f6d(J=1):\SI{20711}{cm^{-1}}$ intermediate state, using laser radiation at 484.3 nm for the first step and 1164.3 nm for the second step excitation. The low angular momenta yield a smaller number of HFS components which facilitates the data evaluation. This experiment was conducted at the Maier-Leibnitz-Laboratorium at LMU Munich using a $^{233}$U source, which produces thorium recoil ions via alpha decay with 2\% of the ions being in the isomeric nuclear state. A detailed description of the experiment and the results of the Doppler-free hyperfine spectroscopy for the nuclear ground and isomeric states are given in Ref. \cite{thielking_laser_2017}.

To measure the frequency detuning of the lasers during the scanning, two temperature stabilized confocal cavities, one for visible and one for infrared light, are used. The infrared cavity is placed in vacuum for higher long-term stability. The transmission signals of the cavities are recorded simultaneously with the fluorescence spectra. For the two-step excitation and the depletion measurement, the frequencies of the 484.3 nm laser and the 459.1 nm laser need to be stabilized. This is achieved by a computer-based locking system using a Fizeau interferometer (HighFinesse WS7).

%
\textit{Results and discussion}

\begin{table*}[htb]
\caption{Magnetic dipole and electric qudrupole hyperfine constants $A$ and $B$ for the ground and four excited states of $^{229}$Th$^{2+}$. The theoretical results, obtained using CI+MBPT and the MCDF method, are compared to the experimental values, if available. \label{tab:doubly 229}}
\begin{tabular}{lcr|ccc|ccc}
\hline
\hline
\multicolumn{3}{c|}{energy level} & \multicolumn{3}{c|}{$A [$MHz$]$} & \multicolumn{3}{c}{$B [$MHz$]$}\\
configuration & $J^{\Pi}$ & energy $[$cm$]^{-1}$ & CI+MBPT & MCDF & exp. & CI+MBPT & MCDF & exp.\\
\hline
$[$Rn$] + 5f6d$ & $4^-$ & $0$     & $64(17)$  & $81(4)$  & $-$      & $3287(630)$ & $3008(260)$   & $-$\\
$[$Rn$] + 6d^2$ & $2^+$ & $63$    & $143(47)$ & $162(8)$ & $151(8)$ & $68(23)$   & $71(7)$   & $73(27)$\\
$[$Rn$] + 5f^2$ & $4^+$ & $15148$ & $38(3)$   & $72(3)$  & $-$      & $1221(390)$ & $1910(200)$  & $-$\\
$[$Rn$] + 5f6d$ & $1^-$ & $20711$ & $109(36)$ & $90(4)$  & $88(5)$  & $839(220)$ & $689(110)$  & $901(18)$\\
$[$Rn$] + 5f^2$ & $4^+$ & $21784$ & $8(36)$   & $26(2)$  & $-$      & $65(21)$   & $39(45)$  & $-$\\
\hline
\hline
\end{tabular}
\end{table*}
\begin{figure}[htb]
\includegraphics[width=\linewidth]{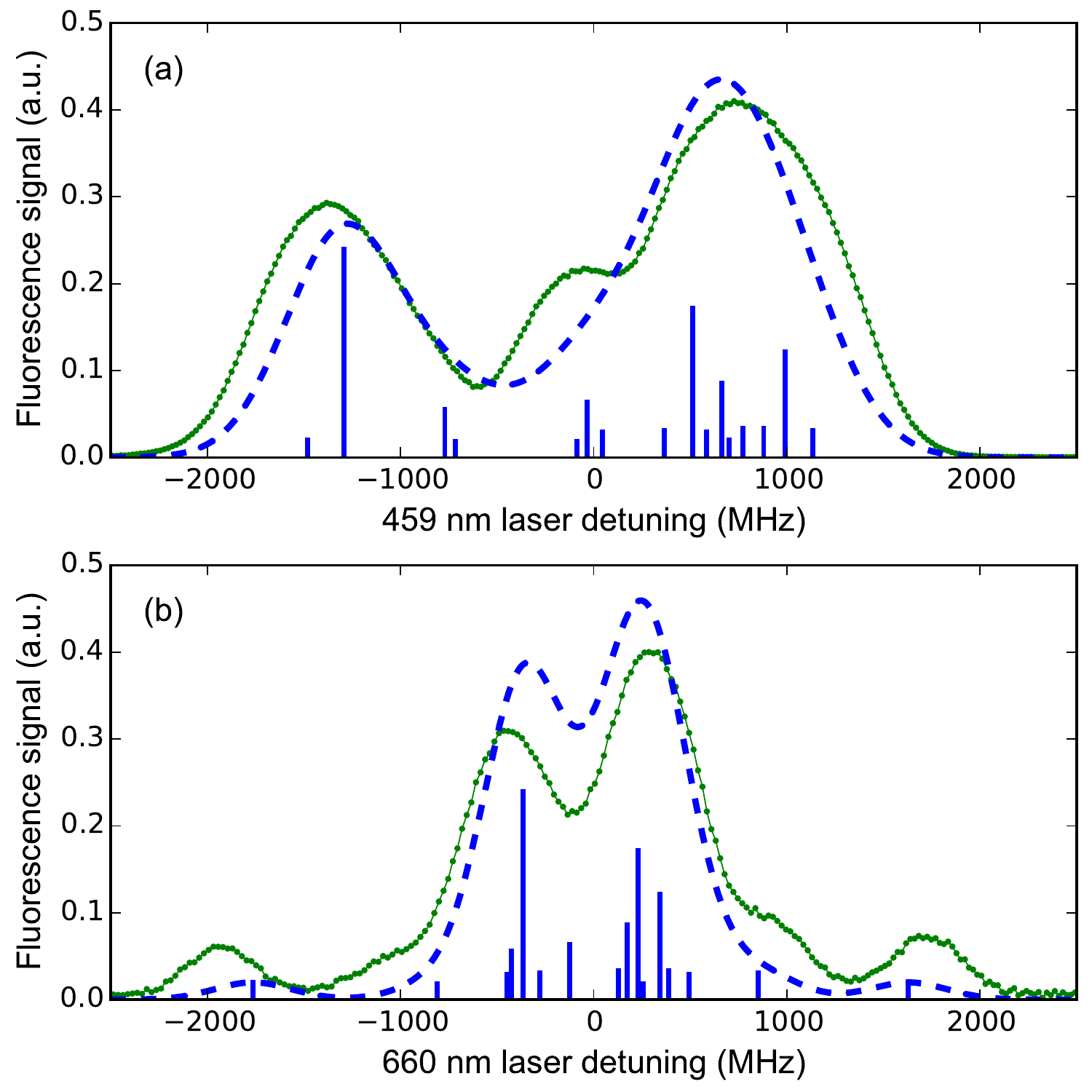}
\caption{\label{fig:spectra} Recorded $^{229}$Th$^{2+}$ hyperfine spectra. a) transition $5f6d(J=4):\SI{0}{cm^{-1}}$ to $5f^2(J=4):\SI{21784}{cm^{-1}}$, b) transition $5f6d(J=4):\SI{0}{cm^{-1}}$ to $5f^2(J=4):\SI{15148}{cm^{-1}}$. Each subfigure shows the experimental data (connected dots) together with the positions of the individual resonances as predicted by MCDF calculations and the resulting theoretical spectrum (dashed line).}
\end{figure}
The hyperfine coefficients for $^{229}$Th$^{3+}$ where measured with high precision \cite{campbell_single-ion_2012} and theoretical calculations of these coefficients have been presented in Ref. \cite{safronova_magnetic_2013}, including the extraction of the values $\mu_{gr} = \SI{0.36}{\mu_N}$ and $Q_{gr}=\SI{3.11}{eb}$ for the moments of the ground state $^{229}$Th nucleus. We used these moments to obtain the hyperfine constants for the ground and the first few excited states of $^{229}$Th$^{3+}$ using the MCDF method. We found these results in good agreement with previous calculations and, thus, the experimental values. Although triply charged thorium is a considerably simpler system, it resembles a very good test case for our calculations for Th$^{2+}$, because in both ions $f$ and $d$ orbitals are populated for the two lowest states, favouring strong core-valence correlation effects. The good agreement with the experiment, thus, gives us confidence that we are able to numerically control these correlations, also in the case of Th$^{2+}$.

While our calculations for triply charged thorium were an important benchmark, the complexity of the problem for doubly charged thorium is still considerably higher. The Th$^{2+}$ ion has $45$ levels in the desired energy range between $\SI{0}{cm^{-1}}$ and $\SI{25000}{cm^{-1}}$, ranging from three ($J^\pi=1^-$) to six ($J^\Pi=2^+$ and $J^\Pi=3^-$) for each pair of $J$ and $\Pi$ [cf. Eq. \eqref{eq:wf ci}]. However only a few levels were adressed in the experiment, all of these levels need to be described to the same accuracy to obtain reliable theoretical results. Therefore the achievable accuracy of our calculations is worse than in other effective two-electron systems.

To obtain results for the level structure of the $^{229}$Th$^{2+}$ ion with the MCDF method we first performed calculations, where the set of CSFs was constructed from a closed radon core allowing for double excitations of the two valence electrons. We added correlation layers until the energies of the calculated levels completely converged. These results were then used to perform calculations, where we subsequently added single excitations from the core up to the argon shell. After also allowing for core-valence excitations from the $6p$ and $5d$ orbitals we achieved a very good convergence of the results eventually. Our agreement with experimental level energies \cite{a._kramida_notitle_2015} is always better than $3\%$, except for the lowest $6d^2(J=2):\SI{63}{cm^{-1}}$ state. The energy of this state is particularly difficult to obtain, because it is almost degenerate with the $5f6d(J=4):\SI{0}{cm^{-1}}$ ground state. The accuracy of the CI+MBPT calculation was somewhat lower. The reason is strong correlations between two valence electrons and the outermost core shells. In order to increase the accuracy, one needs to go beyond the second order MBPT and use CI+AO (CI+all-order) \cite{Koz04, SKJJ09, OMPS15}. Our main goal here is to check that two very different methods give consistent results. For this purpose our CI+MBPT calculation is sufficient.

After we obtained the many-electron wave functions, we calculated the hyperfine coefficients for all states that were investigated experimentally. First we used these results to analyze the spectra that have been obtained using the single step excitation scheme (cf. Fig. \ref{fig:levels}). As seen in Fig. \ref{fig:spectra} the limited resolution of these spectra does not allow to extract experimental values of the hyperfine constants for the electronic ground and the two $5f^2(J=4)$ states. Instead we utilized the MCDF results for the $A$ and $B$ constants of these states (cf. Tab. \ref{tab:doubly 229}) to make a prediction for the positions of the individual resonances and the overall shape of the hyperfine spectrum. In Fig. \ref{fig:spectra} this prediction is shown alongside the experimentally recorded spectra. While the positions of the main resonances match rather well, some discrepancies in the line intensities occur, especially for the $5f6d(J=4):\SI{0}{cm^{-1}}$ to $5f^2(J=4):\SI{21784}{cm^{-1}}$ transition. Moreover the overall width of the calculated spectrum is slightly smaller than it was measured. This suggests that our theory might underestimate the electric quadrupole constant $B$ for the states shown in Fig. \ref{fig:spectra}.

In contrast to the spectra recorded from the single-step excitation which were limited in resolution by Doppler broadening, the Doppler-free spectra of the $6d^2(J=2):\SI{63}{cm^{-1}}$ and $5f6d(J=1):\SI{20711}{cm^{-1}}$ states allowed for a precise extraction of the hyperfine constants $A$ and $B$. In Tab. \ref{tab:doubly 229} we show our numerical results aside with these experimentally extracted values. The theoretical uncertainty of the MCDF results is obtained from a convergence analysis with respect to the addition of correlation layers as well as the number of opened core shells. The uncertainty of the CI+MBPT results is determined by the neglected high-order MBPT terms. We estimated them from the size of the second order MBPT and RPA corrections. We can see in Tab. \ref{tab:doubly 229} that the agreement between theory and experiment is very good for the magnetic hyperfine constant $A$. In the case of the electric quadrupole constant $B$ the agreement is slightly worse but still satisfying.
\begin{table*}[htb]
\caption{Atomic part of the magnetic dipole hyperfine constant $A/\mu_{iso}$ for two excited states of $^{229m}$Th$^{2+}$. The results have been obtained using (a) the CI+MBPT and (b) the MCDF method. The experimental values presented in Ref. \cite{thielking_laser_2017} are utilized to extract the magnetic dipole $\mu_{iso}$ and electric quadrupole $Q_{iso}$ moment of the $^{229m}$Th nuclear isomer. \label{tab:doubly 229m}}
\begin{tabular}{lcr|cc|c|cc|cc|c|cc}
\hline
\hline
\multicolumn{3}{c|}{energy level} & \multicolumn{2}{c|}{$A/\mu_{iso}$ $[\si{MHz\per\mu_N}]$} & $A$ $[\si{MHz}]$ & \multicolumn{2}{c|}{$\mu_{iso}$ $[\si{\mu_N}]$} & \multicolumn{2}{c|}{$B/Q_{iso}$ $[\si{MHz\per eb}]$} & $B$ $[\si{MHz}]$ & \multicolumn{2}{c}{$Q_{iso}$ $[\si{eb}]$} \\
configuration & $J^{\Pi}$ & energy $[\si{cm^{-1}}]$ & (a) & (b) & exp. & (a) & (b) & (a) & (b) & exp. & (a) & (b)\\
\hline
$[$Rn$] + 6d^2$ & $2^+$ & $63$ & $660$ & $750$ & $-263(29)$ & $-0.38$ & $-0.35$ & $22$ & $23$ & $53(65)$ & $-$ & $-$ \\
$[$Rn$] + 5f6d$ & $1^-$ & $20711$ & $506$ & $419$ & $-151(22)$ & $-0.30$ & $-0.36$ & $270$ & $229$ & $498(15)$ & $1.84$ & $2.25$ \\\hline
\hline
\end{tabular}
\end{table*}

To obtain the values shown in Tab. \ref{tab:doubly 229} we used in our calculations of hyperfine constants $A$ and $B$ (cf. Eqs. \eqref{eq:hfs constants}) the values for the nuclear moments of $^{229}$Th from Ref. \cite{safronova_magnetic_2013}. While these values for the ground state nucleus are commonly accepted, the measured value of the nuclear dipole moment of the nuclear isomer $^{229m}$Th \cite{thielking_laser_2017} disagrees strongly with previous theoretical works \cite{dykhne_matrix_1998, litvinova_nuclear_2009}. As an example for a possible application of our calculations and to help to resolve this controversy we calculated the atomic parts $A/\mu_{iso}$ and $B/Q_{iso}$ of the hyperfine constants in the doubly charged nuclear isomer $^{229m}$Th$^{2+}$, which has a nuclear spin of $I_{iso}=\frac{3}{2}$ in contrast to the nuclear ground state with $I_{gr}=\frac{5}{2}$. By combining these calculations with measurements for $A$ and $B$, we can extract the nuclear moments of the nuclear isomer to a high precision. The results of these calculations are shown in Tab. \ref{tab:doubly 229m}. Experimentally $\mu_{iso}$ has been determined to be $\SI{-0.37 \pm 0.06}{\mu_N}$. As seen in the table this coincides with our consideration which gives an average value of $\mu_{iso}=\SI{-0.35}{\mu_N}$. Therefore, our calculations affirm the measurements and thus provide a stronger foundation for the obtained value of the magnetic dipole moment of the nuclear isomer $^{229m}$Th. An analogue consideration for $Q_{iso}$ shows a very good agreement between the value obtained using the CI+MBPT method and the experimental value of $Q_{iso}=\SI{1.74 \pm 0.08}{eb}$, while the result from the MCDF method is slightly above the measured result.

\textit{Concluding remarks}

In summary we aimed for a theoretical prediction of the hyperfine structure of doubly charged thorium with the nucleus being in the ground, and the isomeric state. Therefore we employed two different approaches (i) a combination of many body perturbation theory and configuration interaction and (ii) the multi-configurational Dirac-Fock method. We performed large scale calculations for the hyperfine constants $A$ and $B$ in $^{229}$Th$^{2+}$. The results of these calculations agree excellently with experimental values and thus provide a benchmark for future studies. These studies include a variety of possible excitation processes for the nuclear isomer. A precise knowledge about the hyperfine structure in doubly charged thorium is crucial to confirm whether the excitation was successful. Finally we also obtained the magnetic dipole moment of the nuclear isomer using measurements of the hyperfine constant $A$ and calculations for the ratio $A/\mu_{iso}$. Our result rules out a previous theoretical work that estimated the magnetic dipole moment to be $\mu_{iso}=\SI{-0.076}{\mu_N}$. Therefore our theory allows for a better understanding of the low lying nuclear isomer. Combined with further theoretical and experimental investigations our results will moreover help to formulate new tests of fundamental physics in thorium \cite{litvinova_nuclear_2009, flambaum_enhanced_2006, berengut_proposed_2009}.

\begin{acknowledgments}
RAM and AVM would like to thank Mikhail Kozlov for his support and many helpful discussions. RAM acknowledges support by the RS-APS. AVM acknowledges support by the Ministry of Education and Science of the Russian Federation (Grant No. 3.1463.2017/4.6) and by RFBR (Grant No. 17-02-00216 А) We thank Lars von der Wense, Benedikt Seiferle and Peter Thirolf from LMU Munich for their contributions to joint experiments. We acknowledge financial support from the  European  Union's  Horizon  2020  Research  and  Innovation  Programme  under  Grant  
Agreement No. 664732 (nuClock) and from DFG through CRC 1227 (DQ-mat, project B04). 
\end{acknowledgments}
\bibliography{quellen.bib}
\end{document}